\documentclass[conference]{IEEEtran}
\IEEEoverridecommandlockouts

\usepackage{cite}
\usepackage{amsmath,amssymb,amsfonts}
\usepackage{algorithmic}
\usepackage{graphicx}
\usepackage{textcomp}
\usepackage{xcolor}

\usepackage{amsmath,amsfonts}

\usepackage{algorithm}
\usepackage{array}

\usepackage{stfloats}
\usepackage{url,hyperref}
\usepackage{verbatim}
\usepackage{graphicx}
\usepackage{cite}
\usepackage{comment}
\usepackage{ulem}
\usepackage{tabularray}
\usepackage{subcaption}

\usepackage{multirow}

\def\BibTeX{{\rm B\kern-.05em{\sc i\kern-.025em b}\kern-.08em
    T\kern-.1667em\lower.7ex\hbox{E}\kern-.125emX}}

\DeclareRobustCommand*{\IEEEauthorrefmark}[1]{
    \raisebox{0pt}[0pt][0pt]{\textsuperscript{\footnotesize\ensuremath{#1}}}}

\begin{document}

\title{Orthogonal Disentanglement with Projected Feature Alignment for Multimodal Emotion Recognition in Conversation\\
}

\author{
	\IEEEauthorblockN{
		Xinyi Che\IEEEauthorrefmark{1}, 
		Wenbo Wang\IEEEauthorrefmark{2}, 
		Jian Guan\IEEEauthorrefmark{3}, 
		Qijun Zhao\IEEEauthorrefmark{1}\IEEEauthorrefmark{*}} 
	\IEEEauthorblockA{\IEEEauthorrefmark{1}School of Computer Science\\ Sichuan University, Chengdu 610065, China}
	\IEEEauthorblockA{\IEEEauthorrefmark{2}Faculty of Computing\\ Harbin Institute of Technology, Harbin 150001, China}
    	\IEEEauthorblockA{\IEEEauthorrefmark{3}College of Computer Science and Technology\\ Harbin Engineering University, Harbin 150001, China}
	\IEEEauthorblockA{\IEEEauthorrefmark{1}School of Computer Science\\ Sichuan University, Chengdu 610065, China}
}

\maketitle

\begin{abstract}
Multimodal Emotion Recognition in Conversation (MERC) significantly enhances emotion recognition performance by integrating complementary emotional cues from text, audio, and visual modalities. While existing methods commonly utilize techniques such as contrastive learning and cross-attention mechanisms to align cross-modal emotional semantics, they typically overlook modality-specific emotional nuances like micro-expressions, tone variations, and sarcastic language. To overcome these limitations, we propose Orthogonal Disentanglement with Projected Feature Alignment (OD-PFA), a novel framework designed explicitly to capture both shared semantics and modality-specific emotional cues. Our approach first decouples unimodal features into shared and modality-specific components. An orthogonal disentanglement strategy (OD) enforces effective separation between these components, aided by a reconstruction loss to maintain critical emotional information from each modality. Additionally, a projected feature alignment strategy (PFA) maps shared features across modalities into a common latent space and applies a cross-modal consistency alignment loss to enhance semantic coherence. Extensive evaluations on widely-used benchmark datasets, IEMOCAP and MELD, demonstrate effectiveness of our proposed OD-PFA multimodal in emotion recognition tasks, as compared with the state-of-the-art approaches.
\end{abstract}


\section{Introduction}

Multimodal Emotion Recognition in Conversation (MERC) enhances emotion recognition by integrating complementary emotion-related cues from multiple modalities, including text, audio, and visuals \cite{jing2024dq,wang2024dynamic, ghosal-etal-2019-dialoguegcn, hu-etal-2021-mmgcn, fu2025himul}. 
This multimodal integration enables MERC to capture a wider spectrum of emotional expressions, resulting in significant performance gains over traditional unimodal approaches \cite{jiang2024self,tu2022sentiment,zhang2023cross}.

To effectively integrate these cues from multiple modalities, most existing methods adopt contrastive learning \cite{shou2024adversarial,li2024enhancing} and cross-attention \cite{guo2024speaker,yang2023self,ai2024gcn} to capture their consistent semantics based on their similarity.
For instance, MGCMA \cite{shou2024adversarial} utilizes a distribution-based alignment module to exploit coarse-grained alignment representations among three modalities via contrastive learning. 
SACCMA\cite{guo2024speaker} employs a cross-modal attention mechanism to maximize the consistent information from each modality. 
However, all above methods only focus on exploiting modality invariant information, which can be shared across all modalities, and ignore modality-specific emotion-related cues, such as micro-expressions, tone variations and sarcastic language.

Recently, Joyful \cite{li-etal-2023-joyful} employs a global contextual fusion module to extract global features shared across modalities, along with a specific fusion module to fuse modality-specific cues. However, neither its smoothing strategy nor reconstruction loss can explicitly align consistent semantics for global features. This limits its effectiveness in emotion recognition.

\begin{figure*}[t]
    \centering
    \includegraphics[width=0.90\linewidth]{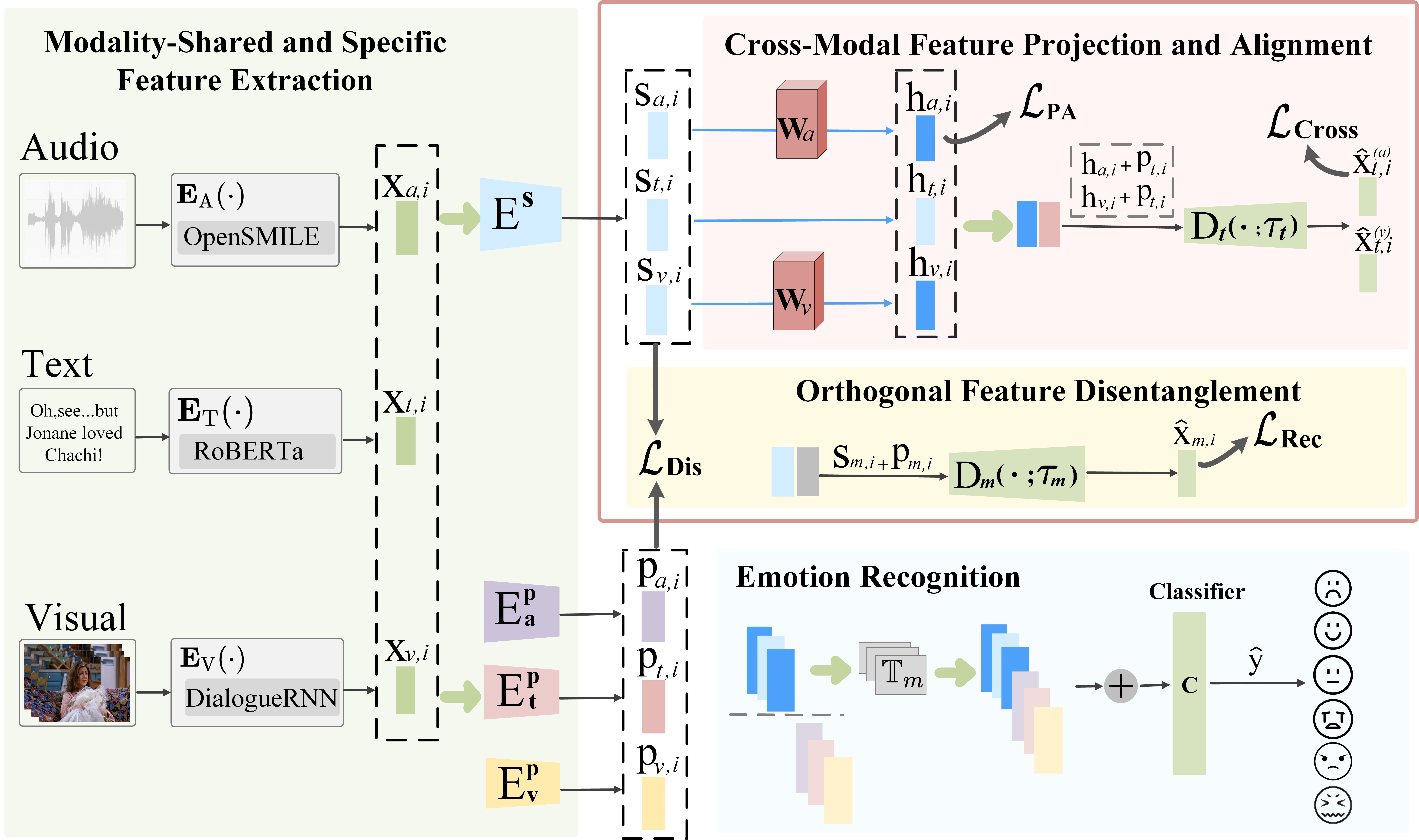}
    \caption{Overall framework of our proposed OD-PFA for multi-modal emotion recognition in conversation. 
    It consists of four components: Multi-Modality Feature Extraction, Orthogonal Feature Disentanglement, Cross-Modal Feature Projection and Alignment, and the Emotion Recognition.} 
    \label{multimodal_2}
\end{figure*}

To this end, we propose an Orthogonal Disentanglement and Projected Feature Alignment (OD-PFA) approach for the MERC task. Specifically, the unimodal features from each modality are first decoupled into shared and modality-specific components, enabling the model to simultaneously capture consistent semantics across all modalities and preserve the distinct cues unique to each modality. After this, we propose an orthogonal disentanglement (OD) strategy combined with a projected feature alignment (PFA) strategy to refine the extracted shared and specific features.
Specifically, the OD strategy applies an orthogonal constraint to enhance the separation between shared and specific features, moreover, a unimodal feature reconstruction loss is incorporated to ensure that critical emotion-related information can be well preserved in both types of features of each modality.
Meanwhile, the PFA strategy employs a feature projection operation to map the shared features of three modalities into a common latent space, and introduces a cross-modal consistency alignment loss to further improve the semantic consistency of these aligned shared features.
Finally, the modality-specific features and the projected shared features are integrated to perform emotion recognition.
Extensive experiments conducted on widely-used IEMOCAP \cite{busso2008iemocap} and MELD \cite{poria-etal-2019-meld} datasets demonstrate effectiveness of our proposed method,  achieving comparable performance to the state-of-the-art approaches.

\section{Proposed Method}

This section presents the details of our proposed OD-PFA for MERC, which consists of Multi-Modality Feature Extraction, Orthogonal Feature Disentanglement, Cross-Modal Feature Projection and Alignment, and the Emotion Recognition. The overall framework is illustrated in Fig. \ref{multimodal_2}.

\subsection{Multi-Modality Feature Extraction}
\subsubsection{Unimodal Feature Extraction}

Following \cite{shi2023multiemo}, for a conversation $\mho$ with $N$ utterances (i.e., $\mho = {\mathcal{U}_{1}, \dots, \mathcal{U}_{n}, \dots, \mathcal{U}_{N}}$), we extract unimodal conversation-level features for each modality. Specifically, audio features $\mathbf{X}_{a}\in \mathbb{R}^{N \times d}$ are obtained using the OpenSMILE toolkit \cite{eyben2010opensmile} and DialogRNN \cite{majumder2019dialoguernn}; text features $\mathbf{X}_{t}\in \mathbb{R}^{N \times d}$ are extracted via a RoBERTa model \cite{kim2021emoberta} followed by a fully connected layer; and visual features $\mathbf{X}_{v}\in \mathbb{R}^{N \times d}$ are derived using VisExtNet \cite{shi2023multiemo} combined with DialogRNN. Here, $d$ denotes the dimensionality of the unimodal features. Based on these, $\mathbf{x}_{m,i} \in \mathbb{R}^d$ represents the unimodal feature of the $i$-th utterance in conversation $\mho$, where $m \in \{a, t, v\}$ indicates the modality.

\subsubsection{Modality-Shared and Specific Feature Extraction}
Here, we leverage a shared encoder, which is shared by all modalities to capture the consistent semantics across modalities for each modality as follows,
\begin{equation}
    \mathbf{s}_{m, i} = \mathbf{E}^{{\text{s}}}(\mathbf{x}_{m, i}; \lambda)
\end{equation}
where \(\mathbf{s}_{m, i} \in \mathbb{R}^d\) denotes the shared feature in modality \(m\), and \(\lambda\) represents the parameters of the shared encoder, which consists of two fully connected layers.

Meanwhile, we also employ three modality-specific encoders to capture specific features individually for each modality, which can preserve the modality-specific semantics, as follows,
\begin{equation}
    \mathbf{p}_{m, i} = \mathbf{E}^{\text{p}}_{m}(\mathbf{x}_{m, i}; \psi_{m})
\end{equation}
where \(\mathbf{p}_{m, i} \in \mathbb{R}^d\) denotes the specific feature of modality \(m\), and \(\psi_m\) refers to the parameters of the corresponding specific encoder,  which also realized via two fully connected layers.

\subsection{Orthogonal Feature Disentanglement }
Our OD strategy $\mathcal{L}_{\text{OD}}$ consists of an orthogonal constraint $\mathcal{L}_{\text{Dis}}$ to promote distinctiveness between shared and specific features, and a unimodal reconstruction loss $\mathcal{L}_{\text{Rec}}$ to preserve essential emotion-related information in both types of features. 
Thus, $\mathcal{L}_{\text{OD}}$ can be denoted as,
\begin{equation}
\mathcal{L}_{\text{OD}} = \alpha \cdot \mathcal{L}_{\text{Dis}} + \beta \cdot \mathcal{L}_{\text{Rec}}
\end{equation}
where $\alpha$ and $\beta$ are hyperparameters.

Inspired by \cite{hazarika2020misa,li2023decoupled}, we apply an orthogonal constraint to minimize the similarity between shared and specific representations, ensuring they capture distinct semantic aspects by enforcing a 90-degree angle between them. This constraint is applied both within the same modality and across different modalities. Therefore, the disentangle loss $\mathcal{L}_{\text{Dis}}$ can be calculated as,

\begin{equation}
\mathcal{L}_{\text{Dis}} = \sum_{m \in \{t, v, a\}} (\mathbf{s}_{m, i}^\top \cdot \mathbf{p}_{m, i})^2 + \sum_{\substack{(m_1, m_2) \in \\ \{(t, a), (t, v), (a, v)\}}} (\mathbf{s}_{m_1, i}^\top \cdot \mathbf{p}_{m_2, i})^2
\end{equation}

To avoid the loss of emotion related information caused by the orthogonal constraint and reduce the risk of learn trivial or uninformative representations, three decoders are utilized to reconstruct the original unimodal feature of each modality based on the decoupled shared and specific features, as follows,
\begin{equation}
    \hat{\mathbf{x}}_{m, i} = \mathbf{D}_{m}(\mathbf{s}_{m, i} + \mathbf{p}_{m, i}; \tau_{m})
\end{equation}
where $\hat{\mathbf{x}}_{m, i}$ denotes the reconstruct unimodal feature of modality $m$, and \(\tau_m\) refers to the parameters of the corresponding decoder, which is realized with two fully connected layers. Thus, the reconstruction loss  $\mathcal{L}_{\text{Rec}}$ is denoted as,
\begin{equation}
        \mathcal{L}_{\text{Rec}} = \frac{1}{3} \sum_{m\in\left \{ t,a,v \right \}}^{} \left \| \mathbf{x}_{m, i}  -\hat{\mathbf{x}}_{m, i} \right \|^{2}_{2}  
\end{equation}

\subsection{Cross-Modal Feature Projection and Alignment}
Here, we propose a cross-modal feature projection and alignment strategy to improve semantic consistency across modalities. Specifically, the shared features are first projected into the common textual space and then aligned. To further enhance cross-modal consistency, the projected shared features and textual specific feature are utilized to reconstruct the original textual unimodal feature.
Thus, our PFA strategy $\mathcal{L}_{\text{PFA}}$ can be denoted as,
\begin{equation}
\mathcal{L}_{\text{PFA}} = \gamma \cdot \mathcal{L}_{\text{PA}} + \xi \cdot \mathcal{L}_{\text{Cross}}
\end{equation}
where $\mathcal{L}_{\text{PA}}$ and $\mathcal{L}_{\text{Cross}}$ denote the common space alignment function and the cross-modal reconstruction function, respectively. $\gamma$ and $\xi$ are hyperparameters.

\subsubsection{Shared Space Feature Projection}
We select the textual modality as the reference and project the shared features of the visual and audio modalities (i.e., ${\mathbf{s}'_{v,i}}$ and ${\mathbf{s}'_{a,i}}$) into the textual space as,
\begin{equation}
    \mathbf{h}_{a, i} = \mathbf{s}_{a, i} \cdot \mathbf{W}_{a}^{\top} ,\quad \mathbf{h}_{v, i} = \mathbf{s}_{v, i}\cdot \mathbf{W}_{v}^{\top} ,\quad \mathbf{h}_{t, i} = \mathbf{s}_{t, i}
\end{equation}
where $\mathbf{W}_{a},\mathbf{W}_{v}\in \mathbb{R}^{d \times d}$ are learnable projection matrices. $\mathbf{h}_{a, i}$ and $\mathbf{h}_{v, i}$ represent the projected shared feature from modality $a$ and $v$, respectively. 
After this, a L1 norm based projection alignment function $\mathcal{L}_{\text{PA}}$ is introduced to explicitly minimize the distance between the projected features (i.e., $\mathbf{h}_{a, i}$ and $\mathbf{h}_{v, i}$ ) and its corresponding textual shared feature (i.e., $\mathbf{s}_{t, i}$), as,
\begin{equation}
    \mathcal{L}_{\text{PA}} = \frac{1}{2} (\left \| \mathbf{h}_{a, i}-\mathbf{s}_{t, i} \right \|_{1} +\left \| \mathbf{h}_{v, i} - \mathbf{s}_{t, i} \right \|_{1})
\end{equation}
where $\left \| \cdot \right \|_{1}$ is L1 norm.

\subsubsection{Cross-Modal Consistency Alignment}
To further enhance the semantic consistency across modalities, we attempt to reconstruct the original textual unimodal feature $\mathbf{x}_{t, i}$ by
leveraging the textual specific decoder (i.e., $\mathbf{D}_{t}(\cdot; \tau_{t})$) based on the specific textual feature (i.e., $\mathbf{p}_{t, i}$) and the projected shared features (i.e., $\mathbf{h}^a_{t, i}$ and $\mathbf{h}^v_{t, i}$) from audio and visual modalities, to improve the semantic consistency of these projected shared features, as,
\begin{equation}
   \hat{\mathbf{x}}_{t,i}^{(a)} = \mathbf{D_{t}}(\mathbf{h}_{a, i}+\mathbf{p}_{t, i}, \tau_{t}), \quad  \hat{\mathbf{x}}_{t, i}^{(v)} = \mathbf{D_{t}}(\mathbf{h}_{v, i}+\mathbf{p}_{t, i}, \tau_{t})
\end{equation}
where $\hat{\mathbf{x}}_{t, i}^{(a)}$ and $\hat{\mathbf{x}}_{t, i}^{(v)}$ represents the reconstructed textual unimodal feature obtained by $\mathbf{h}_{a, i}$ and $\mathbf{h}_{v, i}$, respectively.
Therefore, 
a cross-modal consistency alignment loss $\mathcal{L}_{\text{Cross}}$ can be calculated as:
\begin{equation}
    \mathcal{L}_{\text{Cross}} = \frac{1}{2} (\left \| \hat{\mathbf{x}}_{t, i}^{(a)} - \mathbf{x}_{t, i} \right \|^{2}_{2}+\left \| \hat{\mathbf{x}}_{t, i}^{(v)} -\mathbf{x}_{t, i} \right \| ^{2}_{2} )
\end{equation}

Compared to pairwise alignment of shared features across all three modalities \cite{hazarika2020misa,guo2024speaker,meng2024masked}, our PFA strategy not only reduces alignment complexity, but also narrows modality gaps through projection operation, resulting in more efficient and effective alignment. Additionally, the cross-modal consistency alignment ensures that emotion related information is well-preserved and consistent across all modalities.

\subsection{Emotion Recognition}
\label{sec:sec2.4}
For better recognition, we first enhance the projected shared feature of modality $m$ with contextual semantics by concatenating the projected shared features of all utterances within a conversation, and feeding the sequence into an  $L$-layer Transformer (i.e., $\mathbb{T}_m(\cdot)$) with self-attention mechanism \cite{vaswani2017attention} as follows,

\begin{equation}
\mathbf{H}_m = [\mathbf{h}_{m, 0}, \cdots, \mathbf{h}_{m, i}, \cdots \mathbf{h}_{m, \ell}] \in \mathbb{R}^{\ell\times d}
\end{equation}
\begin{equation}
\mathbf{H}'_m = \mathbb{T}_m(\mathbf{H}_m; \mathbf{\delta}_m)
\end{equation}
\begin{equation}
\mathbf{H}'_m = [\mathbf{h}'_{m, 0}, \cdots, \mathbf{h}'_{m, i}, \cdots \mathbf{h}'_{m, \ell}] \in \mathbb{R}^{\ell\times d}
\end{equation}
where $\mathbf{\delta}_m$ denotes the parameters of the Transformer $\mathbb{T}_m(\cdot)$ for modality $m$. ${\mathbf{h}'_{m,i}}$ represents the contextual enhanced projected shared feature of the $i$-th utterance in modality $m$.

Finally, we integrate all specific features (i.e., $\mathbf{p}_{a, i}, \mathbf{p}_{t, i}, \mathbf{p}_{v, i}$) and contextual enhanced projected features ($\mathbf{h}'_{a, i}, \mathbf{h}'_{t, i}, \mathbf{h}'_{v, i}$) of all three modalities (audio, text and visual) as the input of a classifier $\mathbf{C}(\cdot)$ to give the predict result of the emotion $\hat{y}_{i}$, as follows:
\begin{equation}
    \hat{y} = \mathbf{C}([\mathbf{p}_{a, i}, \mathbf{p}_{v, i}, \mathbf{p}_{t, i}, \mathbf{h}'_{a, i},\mathbf{h}'_{t, i}, \mathbf{h}'_{v, i}])
\end{equation}
where the classifier $\mathbf{C}(\cdot)$ consists of a fully-connected layer and a subsequent 2-layer MLP with a ReLU, as in \cite{shi2023multiemo}.

For model optimization, our OD-PFA is trained through a joint objective function $\mathcal{L}_{Total}$ as follows,
\begin{equation}
\mathcal{L}_{Total} =  \mathcal{L}_{\text{OD}} + \mathcal{L}_{\text{PFA}}+  {\mathcal{L}}_{\text{CE}}(\hat{y}, y)
\end{equation}
where $y$ is the groundtruth emotion label corresponding to $\hat{y}$. ${\mathcal{L}}_{\text{CE}}$ is the cross entropy loss.

\section{Experiments}
\subsection{Experimental Setup}
\noindent\textbf{Dataset:} 
We evaluate our method on two widely-used MERC benchmarks: IEMOCAP \cite{busso2008iemocap} and MELD \cite{poria-etal-2019-meld}. IEMOCAP contains $7,433$ utterances from $151$ dyadic dialogues across five sessions, annotated with six emotions: Happy, Sad, Angry, Excited, Frustrated, and Neutral. 
MELD consists of over $13,000$ utterances from over $1,400$ dialogues in the Friends TV series, labeled with seven emotions. We use the official dataset splits for all experiments.

\noindent\textbf{Evaluation Metrics: } We adopt two widely used evaluation metrics for performance evaluation, including Accuracy (Acc) \cite{yang2023self}\cite{guo2024speaker}  and Weighted F1-score (w-F1)
\cite{hu2022mm} \cite{li-etal-2023-joyful}, following the state-of-the-art methods \cite{meng2024masked,guo2024speaker}.

\noindent\textbf{Settings: } The batch size is set to 32 for both IEMOCAP and MELD. The model is trained for 100 epochs on IEMOCAP and 30 epochs on MELD. The Adam optimizer is adopted with $\beta_{1}=0.9$ and $\beta_{2}=0.99$.

\subsection{Performance Comparison}

We compare OD-PFA with several state-of-the-art methods (i.e., DialogGCN \cite{ghosal-etal-2019-dialoguegcn}, MMGCN \cite{wei2019mmgcn}, MMDFN \cite{hu2022mm} MGLRA\cite{meng2024masked} and Joyful \cite{li-etal-2023-joyful}) on both IEMOCAP and MELD datasets. Results are given in Table \ref{iemocap_table}.

\begin{table*}
\centering
\caption{Performance comparison on IEMOCAP and MELD test dataset under the multimodal setting (text, audio and vision).}
\label{iemocap_table}
\scalebox{1}{
\begin{tblr}{
  width = \linewidth,
  rowsep = 1.2pt,
  colsep = 2.4pt,
  cells = {c},
  cell{1}{1} = {r=2}{},
  cell{1}{2} = {c=8}{0.399\linewidth},
  cell{1}{10} = {c=9}{0.447\linewidth},
  vline{3} = {1}{},
  vline{8,10,17} = {2-14}{},
  hline{1} = {0.4pt}, 
    hline{3,10} = {0.4pt}, 
    hline{11} = {0.4pt}, 
  hline{2} = {2-18}{},
}
\hline
Methods      & IEMOCAP &       &         &       &         &            &       &       & MELD    &          &       &         &       &         &       &       &       \\
            & Happy   & Sad   & Neutral & Angry & Excited & Frustrated & Acc   & w-F1  & Neutral & Surprise & Fear  & Sadness & Joy   & Disgust & Anger & Acc   & w-F1  \\
DiaRNN\cite{majumder2019dialoguernn}  & 32.20    & 80.26 & 57.89   & 62.82 & 73.87   & 59.76      & 63.52 & 62.89 & 76.97   & 47.69    &   -    & 20.41   & 50.92 &  -       & 45.52 & 60.31 & 57.66 \\
MMGCN  \cite{wei2019mmgcn}   & 42.34   & 78.67 & 61.73   & 69.00    & 74.33   & 62.32      & 66.36 & 66.22 & 76.33   & 48.15    &   -    & 26.74   & 53.02 &  -       & 46.09 & 60.42 & 58.31 \\
MMDFN  \cite{hu2022mm}   & 42.22   & 78.98 & 66.42   & \uline{69.77} & 75.56   & 66.33      & 68.21 & 68.18 & 77.76   & 50.69    & -     & 22.93   & 54.78 & -       & 47.82 & 62.49 & 59.46 \\
SCMM \cite{yang2023self}     & 
45.37   & 78.76 & 63.54   & 66.05 & 76.70   & 66.18      & -      & 67.53 &    -     &     -     &    -   &   -      &    -   &  -       &  -     &  -     & 59.44 \\
Joyful  \cite{li-etal-2023-joyful}   & 60.94   & \uline{84.42} & 68.24   & 69.95 & 73.54   & 67.55      & 70.55 & 71.03 & 76.80    & 51.91    & -     & 41.78   & 56.89 & -       & 50.71 & 62.53 & 61.77 \\
SACCMA \cite{guo2024speaker}     & 38.60    & 86.53 & 64.90    & 64.56 & 74.52   & 62.99      &  -     & 67.10  &  -       &   -       &  -     &   -      &  -     &  -       &     -  &   -    & 59.30  \\
MGLRA \cite{meng2024masked}     & \uline{63.50}   & 81.50  & \uline{71.50}    & 61.10  & 76.30    & 67.80      & \uline{71.30}  & 70.10  & 80.80    & 59.50     & -     & 27.80    & 66.50 & -       & 48.04 & 66.40  & 64.90  \\
OD-PFA (Ours)       &  \textbf{66.90}  & 80.09 & \textbf{71.58}   & 64.46& \textbf{79.93}   & \uline{67.75}    &  \textbf{72.09} & \textbf{72.34} & \uline{78.30}   & \uline{58.37}     & \textbf{23.26} & \uline{37.54}   & \uline{63.02} & \textbf{27.37}   & \textbf{55.77} & \textbf{66.97} & \textbf{65.68}
\\
\hline
\end{tblr}}
\end{table*}

It can be seen that our OD-PFA achieves the best performance in both Acc and w-F1 on IEMOCAP and MELD, especially excelling in recognizing Happy, Neutral, and Excited emotions on IEMOCAP, and Fear, Disgust, and Anger on MELD, demonstrating its strong effectiveness for MERC.
Our OD-PFA performs better across both metrics than models like DialogRNN \cite{majumder2019dialoguernn}, SCMM \cite{yang2023self}, and SACCMA \cite{guo2024speaker} that rely solely on cross-attention to capture consistent semantics. This underscores the importance of modeling both modality-shared consistency and modality-specific distinctions.
In addition, OD-PFA surpasses Joyful \cite{li-etal-2023-joyful}, validating the effectiveness of our projected feature alignment strategy for aligning shared representations.

\subsection{Ablation Study}
We analyze the effectiveness of our OD-PFA method by individually ablating the $L_{\text{Dis}}$ for disentangling shared and specific features, the $\mathcal{L}_{\text{PA}}$ for projected feature alignment, and the $\mathcal{L}_{\text{Cross}}$ for cross-modal consistency alignment. 
`w/o' is applied to indicate without specified components. The results are presented in Table~\ref{aa}.

\begin{table}
\centering
 \caption{Validation of $L_{\text{Dis}}$,  $\mathcal{L}{_\text{PA}}$ and  $\mathcal{L}_{\text{Cross}}$ for MERC on both datasets.}
  \label{aa}
  \scalebox{1}{
    \tabcolsep=9pt
\renewcommand\arraystretch{1.2}
    {\begin{tabular}
    { c| c c|c c}
    \hline
    \hline
    \multirow{2}*{Model} & \multicolumn{2}{c|}{IEMOCAP}  & \multicolumn{2}{c}{MELD} \\
    \cline{2-5}

    &Acc & w-F1  &Acc & w-F1\\
   \cline{1-5}
    OD-PFA (w/o - Dis)       & 71.90   & 72.07
                         & 66.36  & 65.53\\   
    OD-PFA (w/o - PA)       & 71.72   & 71.98
                         & 66.90  & 65.58\\
    OD-PFA (w/o - Cross)      & 71.78  & 71.99
                        & 65.90  & 65.50 \\
    \cline{1-5}
    \textbf{OD-PFA}     & \textbf{72.09}   & \textbf{72.34} 
                      & \textbf{66.97}  & \textbf{65.68} \\
    \hline
    \hline
\end{tabular}}}
\end{table}

Removing $L_{\text{Dis}}$ leads to performance drops on both datasets, highlighting its importance in enforcing orthogonality between shared and specific features within and across modalities.
Without $\mathcal{L}_{\text{PA}}$, w-F1 drops to 71.98\% on IEMOCAP and 65.58\% on MELD, indicating the necessity of aligning projected audio and visual shared features with the textual representation. 
Similarly, removing $\mathcal{L}_{\text{Cross}}$ degrades performance, showing its role in preserving semantic integrity and avoiding trivial disentanglement.
The complete OD-PFA model performs best on both datasets, highlighting the combination of orthogonal disentanglement strategy and projection feature alignment strategy.

\subsection{Analysis of Projected Shared Features }

To further evaluate the effectiveness of the projection operation,
we conduct an additional experiment to replace the projected features (i.e., $\mathbf{h}_{a, i}, \mathbf{h}_{t, i}, \mathbf{h}_{v, i}$) with original shared features (i.e., $\mathbf{s}_{a, i}, \mathbf{s}_{t, i}, \mathbf{s}_{v, i}$) in Section \ref{sec:sec2.4}, named the ablated version as OD-PFA (w/o - Pro).
The results are shown in Table \ref{shift}.

\begin{table}
\centering
 \caption{Validation of projected shared feature for MERC on both datasets.}
  \label{shift}
  \scalebox{1}{
    \tabcolsep=9pt
\renewcommand\arraystretch{1.2}
    {\begin{tabular}
    { c| c c|c c}
    \hline
    \hline
    \multirow{2}*{Model} & \multicolumn{2}{c|}{IEMOCAP}  & \multicolumn{2}{c}{MELD} \\
    \cline{2-5}

    &Acc & w-F1  &Acc & w-F1\\
   \cline{1-5}

    OD-PFA (w/o - Pro)       & 71.16  & 71.42
                        & 66.02 & 65.62 \\
    \cline{1-5}
    \textbf{OD-PFA}     & \textbf{72.09}   & \textbf{72.34} 
     & \textbf{66.97} & \textbf{65.68} \\
    \hline
    \hline
   
\end{tabular}}}
\end{table}

It is observed that OD-PFA outperforms OD-PFA (w/o-Pro) across all metrics on both datasets. This confirms that the projected shared feature captures more emotion-related information than the original shared feature, further demonstrating the effectiveness of OD-PFA.

\section{Conclusion}
This paper presented an orthogonal disentanglement with projected feature alignment approach for MERC, which captures both shared semantics and modality-specific emotional cues. An orthogonal disentanglement strategy separates shared and specific components while preserving emotional information via a reconstruction loss. Moreover, our approach  projects shared features into a shared latent space and enforces semantic coherence through a consistency loss with the projected feature alignment strategy. Extensive experiments on the IEMOCAP and MELD datasets demonstrate the effectiveness of OD-PFA, achieving state-of-the-art performance.

\normalem
\bibliographystyle{IEEEtran}
\bibliography{mybib}

\end{document}